\documentclass[aps,prl,twocolumn,superscriptaddress,nofootinbib]{revtex4-1}

\usepackage{mathtools}
\usepackage{lipsum}
\usepackage{amsmath,amssymb}
\usepackage{MnSymbol}
\usepackage{graphicx}
\usepackage{color}
\usepackage{hyperref}
\usepackage{url}
\usepackage{slashed}
\usepackage{slashed,bbm}
\usepackage{graphics,psfrag,epsfig}
\usepackage{dsfont}
\usepackage{enumerate}
\usepackage{pbox}
\usepackage{braket}
\usepackage[retainorgcmds]{IEEEtrantools}
\usepackage{xr}
\usepackage{gensymb}
\usepackage{multibib}
\DeclareMathSizes{10}{10}{6}{4}

\graphicspath{{./figures/}}

\begin{document}

	\title{Second Order Topological Superconductivity in $\pi$-Junction Rashba Layers}
	\author{Yanick Volpez}
	\affiliation{Department of Physics, University of Basel, Klingelbergstrasse 82, CH-4056 Basel, Switzerland}	
	\author{Daniel Loss}
	\affiliation{Department of Physics, University of Basel, Klingelbergstrasse 82, CH-4056 Basel, Switzerland}
	\author{Jelena Klinovaja}
	\affiliation{Department of Physics, University of Basel, Klingelbergstrasse 82, CH-4056 Basel, Switzerland}
\date{\today}	
	
\begin{abstract}
 We consider a Josephson junction bilayer consisting of two tunnel-coupled
 two-dimensional electron gas layers with Rashba spin-orbit interaction, proximitized by a top and bottom $s$-wave superconductor with phase difference $\phi $ close to $\pi$. We show that, in the presence of a finite weak in-plane Zeeman field, the bilayer can be driven into a second order topological superconducting phase, hosting two Majorana corner states (MCSs). If $\phi=\pi$, in a rectangular geometry, these zero-energy bound states are located at two opposite corners determined by the direction of the Zeeman field. If the phase difference $\phi$ deviates from $\pi$ by a critical value, one of the two MCSs gets relocated to an adjacent corner. As the phase difference $\phi$ increases further, the system becomes trivially gapped. The obtained MCSs are robust against static and magnetic disorder. We propose two setups that could realize such a model: one is based on controlling $\phi$ by magnetic flux, the other involves an additional layer of randomly-oriented magnetic impurities responsible for the phase shift of $\pi$ in the proximity-induced superconducting pairing.
\end{abstract}
	\maketitle
{\it Introduction.} Topological insulators and superconductors in $d$ spatial dimensions are gapped fermionic phases with topologically protected gapless states on their $(d-1)$-dimensional boundaries. Among the best-known examples are Majorana bound states in one-dimensional $p$-wave superconductors \cite{Kitaev2001,DasSarma2010,Oreg2010,Alicea2010} as well as two(three)-dimensional topological insulators with an insulating bulk and metallic edges (surfaces) \cite{KaneMele20051,KaneMele20052,Wu2006,Bernevig2006,FuKane2007,Moore2007,Hsieh2008,Hsieh2009,Koenig2007}. The theoretical prediction of such unconventional phases of condensed quantum matter has motivated an enormous amount of experimental and theoretical work \cite{HasanKane2010,Potter2011,Chevallier2012,Klinovaja2012_3,Sticlet2012,Klinovaja2012_2,Halperin2012,NadjPerge2013,Klinovaja2013,Braunecker2013,Vazifeh2013,Das2012,Weithofer2014,Manousakis2017,Mourik2012,Deng2012,Finkc2013,Churchill2013, Albrecht2016,Zhan2018,Zhang2011,Sato2017,Ando2015,Stern2016,Hasan2011}. Later, the concept was generalized to second order or more generally, higher order, topological insulators and superconductors \cite{Benalcazar2017,Benalcazar20172,Geier2018,Benalcazar,song2017,Peng2017,Imhof2017,Schindler2018,Hsu2018,Ezawa2018,Ezawa20182,Ezawa20183,Zhu2018,Wang2018,Zhang2018,Wang20182,Liu2018,Yan2018}. In $d$ spatial dimensions, these are gapped phases with topologically protected gapless states on a $(d-n)$ dimensional boundary, where $n$ is the order of the topological phase. 

\begin{figure}[t]
	\centering
	\includegraphics[width=.8\columnwidth]{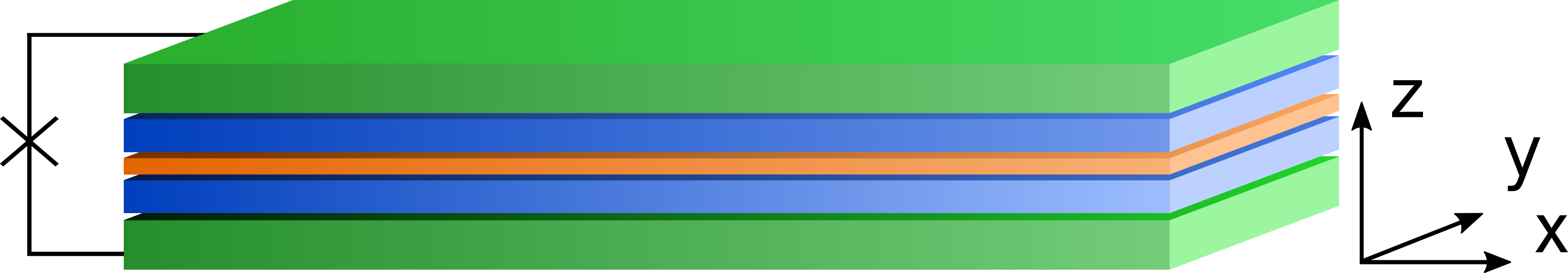}
	\caption{Schematics of the bilayer setup consisting of two 2DEG Rashba layers (blue) separated by a tunnel barrier (orange) and coupled to $s$-wave superconductors (green) that form a Josephson junction. 
The two Rashba SOI vectors are aligned in opposite directions due to the structural asymmetry. 
A magnetic flux ensures a phase difference in the proximity-induced superconductivity between two layers. Alternatively, the phase difference of $\pi$ can be induced by an additional layer of randomly oriented magnetic impurities placed between one of the SC and the Rashba layer.}
	\label{Setup}
\end{figure}

Motivated by these recent developments, we propose a setup that can be controllably brought into the second order topological superconducting (SOTSC) phase. The setup is based on a heterostructure that consists of two two-dimensional electron gas (2DEG) layers with Rashba spin-orbit interaction (SOI) separated by a tunnel barrier. Each of the tunnel-coupled layers is brought into contact with an $s$-wave superconductor (SC) at the top and the bottom to induce the proximity superconductivity, see Fig. \ref{Setup}. Controlling the magnetic flux through the SC loop allows one to control the phase difference $\phi$ between the two parent SCs, and thereby, the phase difference between the proximity gaps in the two layers. Such a setup lies well within experimental reach \cite{Nichele2018,Yacoby2018}. Instead of using a Josephson junction, a second possibility would be to separate one of the Rashba layers from the parent $s$-wave superconductor by an insulating layer of randomly oriented magnetic impurities \cite{Schrade2015}. The phase difference of $\pi$ in the pairing amplitudes arises due to spin-flip tunneling via magnetic impurities \cite{Buzdin,Ryazanov,Spivak,Dam}.

For $\phi=\pi$ and if the tunneling term between the layers dominates over the superconducting pairings, the bilayer is in a helical topological superconducting (HTSC) phase, hosting a Kramers pair of helical edge states counterpropagating along the edges. 
A weak in-plane Zeeman field brings the bilayer into the second order topological superconducting phase. 
The helical edge states are gapped and zero-energy Majorana bound states emerge, in rectangular geometry, at two opposite corners, referred to as Majorana corner states (MCSs). By deviating the phase difference slightly from $\phi=\pi$, the MCSs can be relocated to two adjacent corners. 
The presence of MCSs is robust against potential and magnetic disorder.
 
{\it Model.} The bilayer setup consists of two 2DEGs with strong Rashba SOI proximity coupled to bulk $s$-wave SCs. The top (bottom) layer is labeled by an index $\tau=1$ ($\tau=\bar{1}$). We note that 
in such bilayers the
SOI vectors ${\boldsymbol \alpha}_{\tau}$ are naturally antiparallel due to the structural asymmetry (see Fig. \ref{Setup}) and aligned along the $z$ axis defined to be the normal to the layers \cite{foot}. The Hamiltonian for a Rashba layer reads
\begin{align}\label{HamiltonianFreeLayer}
H_{\tau}= \sum_{\sigma, \sigma'} \int d^2 { k}\  \psi^{\dagger}_{\tau \sigma,\mathbf{k}} \Big[\varepsilon_k-\mu_{\tau} +  \boldsymbol{\alpha}_{\tau}\cdot  \boldsymbol{\sigma} \times {\bf k} \Big]_{\sigma \sigma'}\psi_{\tau \sigma',\mathbf{k}}\, , 
\end{align}
where ${\bf k}=(k_x,k_y)$ is the in-plane momentum, $\varepsilon_k=\hbar^2 k^2/2m$ the kinetic energy, $m$ the effective mass, and $\boldsymbol{\sigma}$ the vector of Pauli matrices acting in spin space. The creation operator $\psi^{\dagger}_{\tau \sigma, {\bf k}}$ acts on an electron with spin projection along $z$ axis $\sigma=\{1,\bar{1}\}$ and in-plane momentum ${\bf k}$ located in layer $\tau= \{1,\bar{1}\}$. The chemical potential $\mu_{\tau}$ is tuned to the SOI energy $E_{\tau,so}= \hbar^2 k_{so,\tau}^2/2m$ with the SOI momentum given by $k_{so,\tau}=m \alpha_{\tau}/\hbar^2$. 
The coupling between the two Rashba layers is described by
\begin{equation}
H_{\Gamma} = \Gamma\sum_{\sigma} \int d^2 { k} \ \Big( \psi^{\dagger}_{1 \sigma, {\bf k}} \psi_{\bar{1} \sigma, {\bf k}} + \text{H.c.} \Big),
\end{equation}
where the spin-conserving tunneling amplitude $\Gamma$ is assumed to be positive. The proximity induced superconductivity is described by
\begin{equation}
H_{\Delta} = \frac{1}{2} \sum_{{\tau,  \sigma, \sigma'}} \int d^2 {k} \ \Big( \Delta_{\tau} \psi^{\dagger}_{\tau \sigma,{\bf k}} [i \sigma_2]_{\sigma \sigma'} \psi^{\dagger}_{\tau \sigma',-{\bf k}} + \text{H.c.}\Big), \label{HamiltonianSC}
\end{equation}
where we assume the pairing amplitude $\Delta_1$ in the top layer to be real and positive, while the pairing amplitude in the second layer can be complex $\Delta_{\bar{1}}=|\Delta_{\bar{1}}| e^{i \phi}$. There are several mechanisms that can produce a phase difference $\phi$. A superconducting loop connecting two SCs enclosing a magnetic flux $\phi$ (see Fig. \ref{Setup}) forms a Josephson junction \cite{Keselman2013,Pientka2017,Nichele2018,Yacoby2018}, that allows one to tune the superconducting phase difference $\phi$ between the two Rashba layers. Alternatively, such a setup could be realized by involving an insulating layer of random magnetic impurities \cite{Schrade2015} between a SC and one of the Rashba layers, such that the superconducting phase difference between two Rashba layers is equal to $\pi$ and does not require any fine-tuning necessary in the first setup.

We also account for the presence of an in-plane Zeeman field of strength $\Delta_Z$ applied along the unit vector $\boldsymbol{n}=(\cos \theta, \sin \theta,0)$. The corresponding term in the Hamiltonian is given by
\begin{equation}\label{HZeeman}
H_{Z} = \Delta_Z \sum_{{\tau, \sigma, \sigma'}} \int d^2 {k} \ \psi^{\dagger}_{\tau \sigma, {\bf k}} [\boldsymbol{n} \cdot \boldsymbol{\sigma}]_{\sigma \sigma'} \psi_{\tau \sigma', {\bf k}}\, .
\end{equation}
The total Hamiltonian reads $H=H_1+H_{\bar{1}} +H_{\Gamma} + H_{\Delta} +H_Z \equiv H_0+H_Z$. Next, we solve $H$ numerically for two geometries. ($i$) Semi-infinite geometry: the system is translationally invariant along the $x$ direction (with momentum $k_x$ as good quantum number) and finite along the $y$ direction of length $L_y$. ($ii$) Finite geometry: the system is finite in both, the $x$ and $y$, directions and of size $L_x \times L_y$, where $L_x$ is the length of the system in the $x$ direction. In what follows, we present the key results of the numerical diagonalization of $H$, while the specific form of the discretized Hamiltonians can be found in the Supplemental Material (SM) \cite{SM}.

\begin{figure}
	\includegraphics[width=.75\columnwidth]{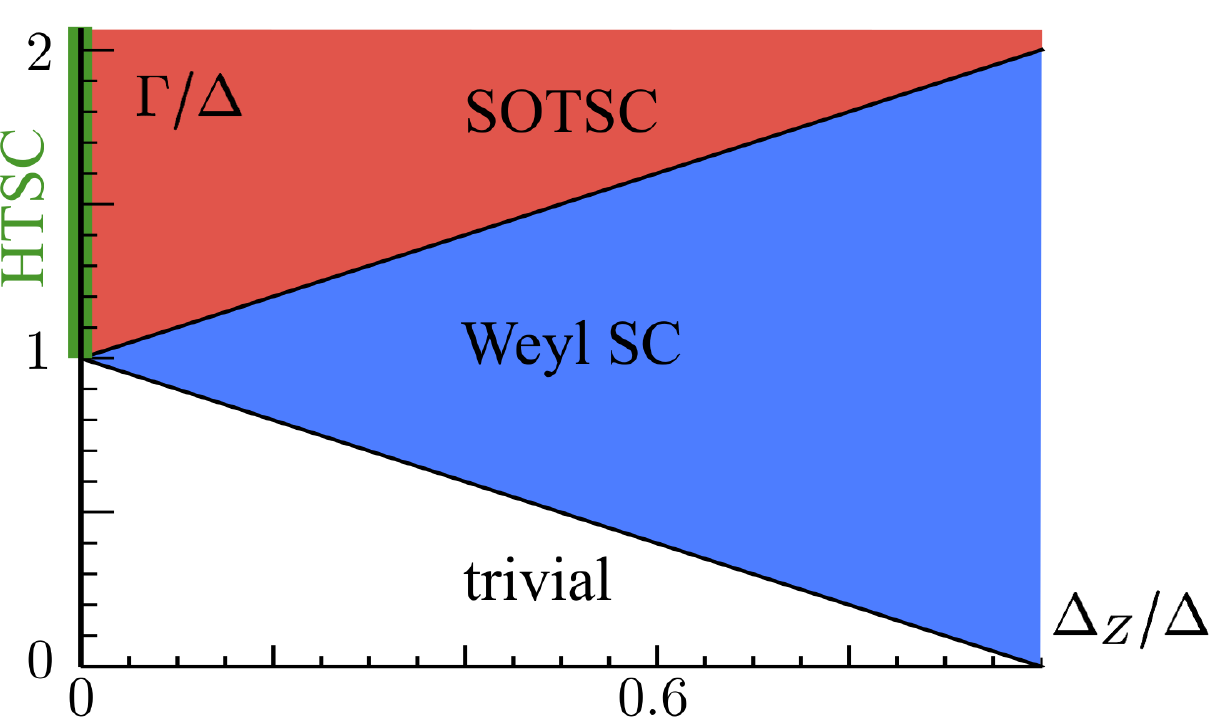}
	\caption{Topological phase diagram of the $\pi$-junction bilayer as function of tunnel amplitude $\Gamma$ and Zeeman energy $\Delta_Z$. Topological phase transitions occur at $\Gamma=|\Delta_Z \pm \Delta|$ (black lines). For $\Delta_Z=0$ and $\Gamma>\Delta$, the system is in the HTSC phase (green line) hosting a Kramers pair of helical edge states. For $\Delta_Z<\Delta$ and $\Gamma>|\Delta+\Delta_Z|$, the system is in the SOTSC 
	phase (red) hosting at two of the corners each a MCS. For $|\Delta_Z-\Delta|<\Gamma<|\Delta_Z+\Delta|$, the system is in the Weyl SC phase (blue) hosting flat zero-energy edge states. Otherwise, the system is in the trivial phase (white). }
	\label{PhaseDiag}
\end{figure}

\begin{figure}[t]
	\centering
	\includegraphics[width=\columnwidth]{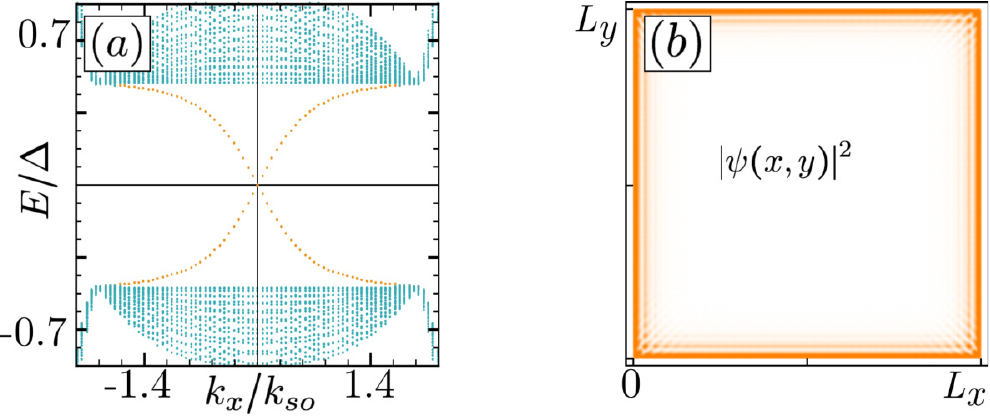}
	\caption{(a) Energy spectrum in the HTSC phase ($\Gamma>\Delta$) for semi-infinite geometry, obtained by exact diagonalization. The bulk states (green dots) are gapped, while both edges host a Kramers pair of dispersive helical edge states (orange dots). (b) Color plot of the probability density $|\psi (x,y)|^2$ of a helical edge state in the finite geometry. The states are exponentially localized at the edge (yellow strips). Numerical parameters: $ k_{so }L_x =k_{so} L_y \approx 860$, $\mu=0$, $\Delta/E_{so} \approx 0.2$, $\phi=\pi$, and $\Gamma/E_{so}\approx 0.6$.}
	\label{HelicalEdgeStates}
\end{figure}

{\it Topological phase diagram of $\pi$-Junction.} If $\Delta_Z=0$ and $\phi=\pi$, the system is in the DIII symmetry class \cite{Ryu2010}. The effective time-reversal (particle-hole) symmetry operator is given by $\Theta=i\sigma_2 \mathcal{K}$ ($\mathcal{P}=\eta_1 \mathcal{K}$), where $\eta_i$ are the Pauli matrices acting in particle-hole space and $\mathcal{K}$ is the complex conjugation operator. 
The corresponding bulk spectrum of $H_0$ is given by 
\begin{align}
&E^2_{\pm}(k) = \varepsilon_k^2+\Delta^2+\Gamma^2+(\alpha k)^2 
 \pm {2}\sqrt{\varepsilon_k^2[\Gamma^2+(\alpha k)^2]+\Delta^2 \Gamma^2  }\,\,.
\end{align}
For simplicity we set $\Delta_1=|\Delta_{\bar{1}}|=\Delta>0$ and $|\boldsymbol{\alpha}_1|=|\boldsymbol{\alpha}_{\bar{1}}|=\alpha$, however, we note that our results remain valid in the more general case as can be easily checked numerically. The bulk gap closes at $k=0$ for $\Gamma= \Delta$. For $\Gamma=0$, the system consists of uncoupled layers, and, thus, is in the trivial phase for $\Gamma<\Delta$, see Fig. \ref{PhaseDiag}. In the regime $\Gamma>\Delta$, the system has a Kramers pair of gapless helical edge states on both boundaries ($y=0$ and $y=L_y$) in the semi-infinite geometry and on each boundary in the finite geometry, see Fig. \ref{HelicalEdgeStates}. These states are exponentially localized at the boundary and have a linear dispersion around $k_x=0$.

Zeeman fields break time-reversal symmetry such that the Hamiltonian $H$ is placed now in the $D$ symmetry class \cite{Ryu2010}. For strong Zeeman field, $\Delta_Z>\Delta$, superconductivity is suppressed such that the system becomes gapless with the gap closing at finite momenta.
At smaller fields, the bulk gap closes at $k=0$ for $\Gamma=|\Delta_Z \pm \Delta|$, indicating a topological phase transition. Obviously, at weak fields, $\Gamma<\Delta-\Delta_Z$, the system stays trivial.
If $\Delta-\Delta_Z<\Gamma<\Delta+\Delta_Z$, the system is in a 2D Weyl SC phase \cite{Tanaka20101,Sato20101,Sato20111,Schnyder20111,Meng20121,Wong20131,Deng20141,Schnyder20151,Hao20171} (see Fig. \ref{PhaseDiag}), with two Weyl cones emerging in the spectrum in direction orthogonal to the Zeeman field. The nodes of these two Weyl cones are connected in momentum space by a zero-energy line describing localized dispersionless edge states. The zero-energy edge states occur only at edges where the nodes of the Weyl cones are not projected onto the same point \cite{SM}.

\begin{figure*}
	\includegraphics[width=.95\textwidth]{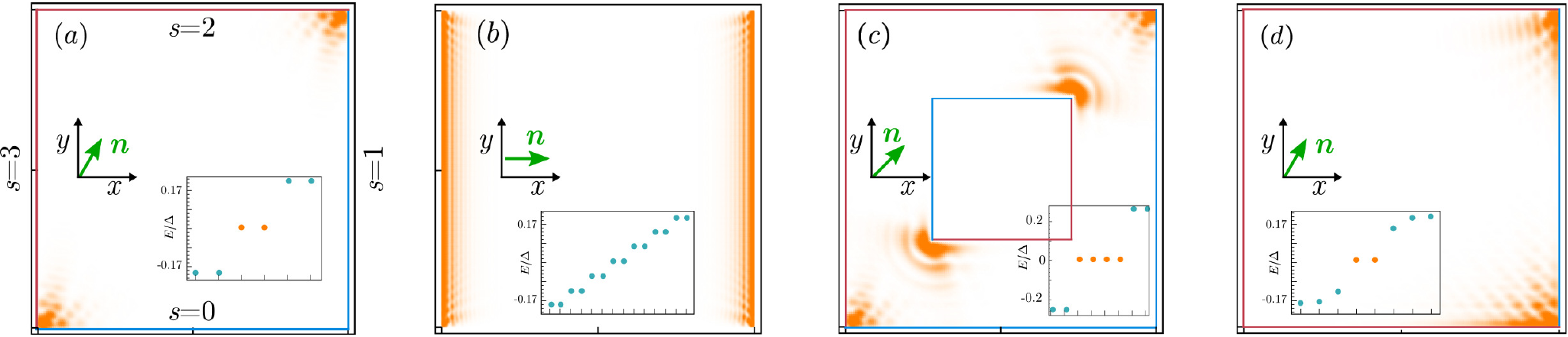}
	\caption{Color plots of probability density and spectrum (insets) of low-energy states in SOTSC phase.
	(a) If the field direction ${\bf n}$ has projection on both edges as well as $|\delta \phi|< \min \{\delta \phi_{c, \parallel}, \delta \phi_{c, \perp} \}$ (here, $\theta=\pi/3$ and $\delta \phi=0$), the zero-energy MCSs (yellow dots in inset) are separated from bulk states (blue dots in inset) by the gap and are located at two opposite corners. For $0<\theta<\pi/2$, the sign of the gap $\Delta_s$ on the $s=0,1$ ($s=2,3$) edge is positive (negative) and shown in blue (red): thus, the lower left (upper right) corner acts as domain wall at which MCSs are localized. (b) If $|\delta \phi|= \delta \phi_{c, \parallel/\perp}$ (here, $\theta=0$ and $\delta \phi=\phi_{c, \perp}=0$), the helical edge states propagating along the corresponding edge stays gapless. The other edges are still gapped. (c) A trivial square region in the center (of the size $L_x/2 \times L_x/2$) leads to two additional MCSs localized at the inner corners. (d) If $\min \{\delta \phi_{c, \parallel}, \delta \phi_{c, \perp} \}< |\delta \phi|< \max \{ \delta \phi_{c, \parallel}, \delta \phi_{c, \perp}\}$ (here $\theta=\pi/3$ and $\delta \phi \approx \pi/3$), only the sign of the gap on the $s=1$ edge is positive. As a result, the two MCSs are located now at two neighboring corners. Numerical parameters are the same as in Fig. \ref{HelicalEdgeStates}, except $k_{so}L_x= k_{so} L_y=430$ and $\Delta_Z/E_{so} = 0.15$.}
	\label{CornerStates}
\end{figure*}

Next, we focus on the SOTSC phase, $\Gamma>\Delta+\Delta_Z$. Generally, finite $\Delta_Z$ opens a gap in the spectrum of the helical edge states of the HTSC phase. The size of the gap depends on the field direction ${\bf n}$ \cite{Volpez2018} (see Fig. \ref{CornerStates}).
For simplicity, we consider samples of rectangular geometry. For fields not parallel to the sample edges, the helical edge states are gapped out. More importantly, two non-degenerate bound states emerge at zero-energy which we can identify as Majorana corner states, each localized at opposite corners, see Fig. \ref{CornerStates}. The localization length of such MCSs along the edges is inversely proportional to the gap. If the system possesses, in addition, mirror symmetry [$\theta=(2p+1) \pi/4$], the localization lengths along $x$ and $y$ axis are the same. As the field deviates from these directions, the localization length along the axis with the smallest gap increases [see Fig. \ref{CornerStates}a] up to the point where the two edges are no longer gapped as the field is aligned along one of edges [see Fig. \ref{CornerStates}b]. If the field is rotated further, the MCSs re-emerge at the other two opposite corners.

Generally, the existence of the two MCSs does not rely on symmetries of the square lattice, nor on the particular shape of the boundary. If the system {\it e.g.} has a circular shape, the two MCSs are localized at the two opposite points where the Zeeman vector ${\bf n}$ crosses the edge. We also checked numerically that MCSs are robust against moderate potential and magnetic disorder \cite{SM}. We can also easily create more than two MCSs by modifying the topology of our setup. For instance, we can allow for a region inside the system to be in the trivial phase, giving rise to an inner boundary at the interface with corresponding edge states.
Such a region could be fabricated by covering, say, the top Rashba layer with a SC layer with {\it e.g.} a rectangular hole. Without Zeeman fields, helical edge states propagate along the outer and inner edges, while for $\Delta_Z>0$, these states get gapped and four MCSs emerge, two at the outer and two at the inner corners, see Fig. \ref{CornerStates}c. 

{\it Analytical treatment and stability of MCSs.} In order to obtain a deeper understanding for the appearance of the MCSs at the two particular corners, we treat the problem analytically. We assume that the system size is large enough to treat all four edges, away from the corners, independently. For simplicity, we focus on rectangular geometries and label the edges counterclockwise by $s=0,1,2,3$ with $s=0$ being the horizontal edge at $y=0$ (see Fig.~\ref{CornerStates}). The effective Hamiltonian describing the helical edge states is given by
\begin{align}\label{HmassTerm}
\mathcal{H}^{(s)}_{eff} &= \hbar v k^{(s)} \beta_3-\Delta_s \beta_2,
\end{align}
where $v$ is the effective Fermi velocity of the edge states, $k^{(s)}$ the momentum along the $s$th edge, $\beta_i$ the Pauli matrices acting on the low-energy subspace, and $\Delta_s$ is the gap opened on the $s$th edge if the time-reversal symmetry is broken. We also note that the Hamiltonian without Zeeman field, $H_0$ [see Eqs. \eqref{HamiltonianFreeLayer}-\eqref{HamiltonianSC}], has rotational symmetry around the $z$-axis, $[H_0,U_{\varphi}]=0$, where $U_{\varphi}=e^{i \varphi \eta_3 J_z/\hbar}$. Here, $J_z=L_z+S_z$ is the $z$ component of the total angular momentum composed of spin $S_z=\hbar \sigma_3/2$ and orbital angular momentum $L_z=-i \hbar (x \partial_y - y \partial_x)$. At the edges, this symmetry reduces to a four-fold rotational symmetry with $\varphi_s = s \pi/2$. Thus, the states localized at the $s$th edge are related to those at $s=0$ by the unitary $U_{\varphi_s}$. For example, $\Phi^{(s)}_{\pm }=U_{\varphi_s} \Phi^{(0)}_{\pm }$, where $\Phi^{(s)}_{\pm }$ are the $s$th edge state wavefunctions at $k^{(s)}=0$. This rotational symmetry ensures that the effective Fermi velocity is the same at all edges. 

So far, we focused on the phase difference being tuned to $\phi=\pi$. This is important for observing the helical edge states protected by time-reversal symmetry but not necessary for observing the MCSs. The SOTSC phase and MCSs are robust against small deviations of the phase difference $\phi=\pi+\delta \phi$, which is of great importance for the scenario in which this phase is tuned by magnetic fluxes. For finite $\delta \phi$ ($|\delta \phi| \ll 1$), the bulk spectrum stays almost the same, while the helical edge states become gapped at $k^{(s)}=0$. The size of the gap for $\delta \phi \neq 0$ can be determined in first order perturbation theory by calculating the expectation value of the corresponding part of $H_{\Delta}$, $\left<\Phi^{(s)}_{+}|\delta H_\Delta |\Phi^{(s)}_{-}\right>$. The gap $\Delta \sin( \phi)/2 \approx -\Delta \delta \phi /2$ is the same on all edges due to rotation symmetry, see SM~\cite{SM}.

As discussed above, a Zeeman field also opens a gap in the spectrum of edge states at $k^{(s)}=0$. 
Since the edge states at $s=0$ have a well-defined spin projection along $y$, only the $x$-component of the field can lead to a gap given by $\left<\Phi^{(0)}_{+}| {\cal H}_Z |\Phi^{(0)}_{-}\right> = \Delta_Z \cos \theta$. The Zeeman term ${\cal H}_Z$ defined in Eq. (\ref{HZeeman}) is not invariant under rotations, $U^{\dagger}_{\varphi_s} {\cal H}_Z U_{\varphi_s} =\Delta_Z[\cos(\theta-\varphi_s) \ \eta_3 \sigma_1 + \sin(\theta-\varphi_s) \sigma_2]/2$. Thus, the gap $\left<\Phi^{(0)}_{1}| U^{\dagger}_{\varphi_s} {\cal H}_Z U_{\varphi_s}|\Phi^{(0)}_{\bar 1}\right>$ opened by ${\cal H}_Z$ is different at different edges and is given by $\Delta_Z \cos(\theta- \varphi_s)$, see SM~\cite{SM}.

Combining the two complementary mechanisms gapping the helical edge states, we find that the gap on the $s$th edge is given by $\Delta_s= \Delta \delta \phi/2+\Delta_Z \cos(\theta- \varphi_s)$. The edges are gapped except if $\delta \phi=\pm\delta \phi_{c, \parallel/\perp}$, where the gap closes at the horizontal ($\delta \phi_{c, \parallel}= 2 \Delta_Z |\cos \theta| / \Delta$) or vertical ($\delta \phi_{c, \perp} = 2 \Delta_Z |\sin \theta|/\Delta$) edges. The effective $\mathcal{H}^{(s)}_{eff}$ brings us back to the first topological models of Jackiw-Rebbi type \cite{Jackiw1976,Jackiw1981}. The mass term $\Delta_s$, opening the gap, can change its sign at a corner where two edges meet. As a result, there is a zero-energy bound state localized at this effective domain wall, which we identify as a MCS. Importantly, the details of the parameter dependence at the domain wall is not important and only the asymptotics away from it matters. This allows us to rely on the effective $\mathcal{H}^{(s)}_{eff}$ derived away from the corners.

There are three regimes to consider. If $|\delta \phi|> \max \{ \delta \phi_{c, \parallel}, \delta \phi_{c, \perp}\}$, the mass term $\Delta_s$ is either positive or negative on all edges, since the uniform gap opened by $\delta \phi$ dominates. In this case, there are no domain walls and thus no MCSs. If $|\delta \phi| < \min \{ \delta \phi_{c, \parallel}, \delta \phi_{c, \perp}\}$, the gap due to the Zeeman field dominates and there are two domain walls that lie on opposite corners along the diagonal, see Fig. \ref{CornerStates}a. In the intermediate regime, $\min \{ \delta \phi_{c, \parallel}, \delta \phi_{c, \perp}\}<|\delta \phi| < \max \{ \delta \phi_{c, \parallel}, \delta \phi_{c, \perp}\}$, the domain walls, and as result the MCSs, are located on neighboring corners, see Fig. \ref{CornerStates}d. Thus, the MCSs exist for a wide range of $\delta \phi$ and their location is governed by both the Zeeman field and $\delta \phi$. While the analytical treatment was done perturbatively, the conditions on the existence of MCSs as well as their location can be confirmed numerically well beyond the perturbative regime, see Fig.~\ref{CornerStates}.

{\it Conclusions.} We studied a bilayer Josephson junction consisting of two tunnel coupled 2DEGs with opposite Rashba SOI and proximity-induced superconducting pairing amplitudes that have a phase difference $\phi \approx \pi$. Alternatively, since the spectrum of the uncoupled layers around $k=0$ is essential for the topological properties of the system, instead of two Rashba layers with opposite SOI separated by a tunneling barrier, one could also use a thin film of a 3D topological insulator \cite{BlackSchaffer2016}. There, the surface states on opposite surfaces have opposite helicity and therefore have the same spectrum around $k=0$ as in the Rashba bilayer setup.
If tunneling between layers dominates over the superconducting pairings and a weak in-plane Zeeman field is applied, the system is in a SOTSC phase hosting two MCSs that are robust against both potential and magnetic disorder.

\emph{Acknowledgments.} This work was supported by the Swiss National Science Foundation and NCCR QSIT. This project received funding from the European Union's Horizon 2020 research and innovation program (ERC Starting Grant, grant agreement No 757725).

\bibliographystyle{unsrt}

\onecolumngrid
\newpage
\vspace*{1cm}
\begin{center}
	\large{\bf Supplemental Material to `Second Order Topological Superconductivity in $\pi$-Junction Rashba Layers' \\}
\end{center}
\begin{center}
	Yanick Volpez, Daniel Loss, and Jelena Klinovaja\\
	{\it Department of Physics, University of Basel, Klingelbergstrasse 82, CH-4056 Basel, Switzerland}
\end{center}
\vspace*{1cm}
\onecolumngrid
\setcounter{equation}{0}
\setcounter{figure}{0}

\section{Appendix A: Discretized lattice models}\label{TBM}

In the main text we present our numerical results for the semi-infinite and the finite geometries in which the Rashba bilayer system is assumed to have a rectangular shape. In this section we explicitly discretize the total Hamiltonian $H$ defined by Eqs. (1)-(4) of the main text.

\subsection*{\it Semi-infinite geometry}

In the semi-infinite geometry, we assume, without loss of generality, that the system is translationally invariant along the $x$ and finite along the $y$ direction with the length $L_y=(N_y-1)a$, where $N_y$ is the number of lattice sites in $y$-direction and $a$ the lattice constant. The total Hamiltonian for the semi-infinite geometry is given by $\bar{H}'=\sum_{k_x}[\bar{H}'_1(k_x)+\bar{H}'_{\bar{1}}(k_x)+\bar{H}'_{\Gamma}(k_x)+\bar{H}'_D(k_x)+	\bar{H}'_{Z}(k_x) ]$ with
\begin{align}
\bar{H}'_{\tau}(k_x) &=\sum_{m} \Big\{ \sum_{\sigma} \Big(-t c^{\dagger}_{k_x \tau (m+1) \sigma} c_{k_x\tau m \sigma} + [-t \cos(k_x a) +\mu_{\tau}/2+2t] c^{\dagger}_{k_x \tau m \sigma} c_{k_x\tau m \sigma} + \text{H.c.} \Big) \nonumber \\
&\hspace{.5cm}+ \tau \tilde{\alpha}\Big[ i(c^{\dagger}_{k_x \tau (m+1) \uparrow}c_{k_x\tau m \downarrow} - c^{\dagger}_{k_x \tau (m-1) \uparrow} c_{k_x\tau m \downarrow})  + 2i \sin(k_x a) c^{\dagger}_{k_x \tau m \uparrow} c_{k_x\tau m \downarrow} + \text{H.c.} \Big] \Big\}, \nonumber \\
\bar{H}'_{\Gamma}(k_x) &= \Gamma \sum_{\sigma} \sum_m \Big( c^{\dagger}_{k_x 1 \sigma m}c_{k_x \bar{1} \sigma m} + \text{H.c.}\Big), \nonumber \\
\bar{H}'_D(k_x) &= \frac{1}{2} \sum_{\tau,m} \sum_{\sigma, \sigma'} \Big(\Delta_{\tau} c^{\dagger}_{k_x \tau m \sigma} [i \sigma_2]_{\sigma \sigma'} c^{\dagger}_{-k_x \tau m \sigma'} + \text{H.c.}\Big), \\
\bar{H}'_{Z}(k_x) &= \Delta_Z \sum_{\tau, m} \sum_{\sigma, \sigma'} c^{\dagger}_{k_x \tau m \sigma} [{\bf n}\cdot \boldsymbol{\sigma}]_{\sigma \sigma'} c_{k_x \tau m \sigma'}  
\end{align}
The operator $c^{\dagger}_{k_x \tau \sigma m}$ creates an electron with momentum $k_x$ and spin projection $\sigma$ (along $z$-axis) in the layer $\tau$ at the lattice site $m$. Here, $t$ is the amplitude for a hopping process between two neighboring lattice sites used to set the effective mass as $t=\hbar^2/(2m a^2)$.
The spin-flip hopping amplitude $\tilde{\alpha}$ is related to the SOI parameter by $\tilde{\alpha}=\alpha/2a$. The spin-orbit energy $E_{so}$ is given by $E_{so} = \tilde{\alpha}^2/t$ \cite{Diego, SOI}.

\subsection*{\it Finite rectangular geometry}

In the finite geometry we assume the system to be finite in both $x$ and $y$ directions and of size $L_x \times L_y=(N_x-1)(N_y-1) a^2$. The indices $(n,m)$ label sites in the two-dimensional square lattice with $n \in \{1,\dots,N_x\}$ and $m \in \{1,\dots, N_y\}$. The total Hamiltonian $\bar{H}=\bar{H}_1+\bar{H}_{\bar{1}}+\bar{H}_{\Gamma}+\bar{H}_{D} +\bar{H}_{Z}$ in the finite geometry is then given by
\begin{align}\label{HamiltonianTB}
\bar{H}_{\tau} &= \sum_{n,m} \Big\{ \sum_{\sigma} \Big(-t_x c^{\dagger}_{\tau \sigma (n+1) m} c_{\tau \sigma n m} -t_y c^{\dagger}_{\tau \sigma n (m+1)} c_{\tau \sigma n m} + \frac{\mu_{\tau}+4t}{2} c^{\dagger}_{\tau \sigma n m} c_{\tau \sigma n m} + \text{H.c.} \Big) \nonumber \\ 
&\hspace{.5cm} + \tau \tilde{\alpha}\Big[ i(c^{\dagger}_{\tau \uparrow n (m+1) }c_{\tau \downarrow n m} - c^{\dagger}_{\tau \uparrow n (m-1)} c_{\tau \downarrow n m}) -(c^{\dagger}_{\tau \uparrow (n+1) m}c_{\tau \downarrow n m} - c^{\dagger}_{\tau \uparrow (n-1) m} c_{\tau \downarrow n m}) + \text{H.c.} \Big] \Big\}, \nonumber \\
\bar{H}_{\Gamma} &= \Gamma \sum_{\substack{\sigma \\ n,m}} \Big( c^{\dagger}_{1 \sigma n m} c_{\bar{1} \sigma n m} + \text{H.c.}\Big), \nonumber \\
\bar{H}_{\Delta} &= \frac{1}{2} \sum_{\substack{\tau \\ n,m}} \sum_{\sigma, \sigma'} \Big(  \Delta_{\tau} c^{\dagger}_{\tau \sigma n m} [i \sigma_2]_{\sigma \sigma'} c^{\dagger}_{\tau \sigma' n m} + \text{H.c.}\Big), \\
\bar{H}_{Z} &= \Delta_Z \sum_{\substack{\tau\\ n,m}} \sum_{\sigma, \sigma'} c^{\dagger}_{ \tau \sigma n m} [{\bf n}\cdot \boldsymbol{\sigma}]_{\sigma \sigma'} c_{ \tau \sigma' n m}  
\end{align}
The operator $c^{\dagger}_{ \tau \sigma n m}$ creates an electron with spin projection $\sigma$ in the layer $\tau$ at the lattice site $(n,m)$. Note that $\bar{H}'$ in the semi-infinite geometry can be obtained from $\bar{H}$ by applying the Fourier transformation in $x$ direction.

\section{Appendix B: 2D Weyl superconductor}

\begin{figure}[b]
	\centering
	\includegraphics[width=.35\textwidth]{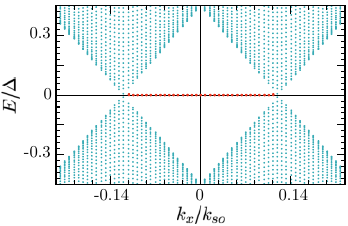}
	\caption{The energy spectrum in the 2D Weyl SC phase in the semi-infinite geometry. The Zeeman field is applied along the $y$ direction such that the Weyl nodes are located at the $k_x$ axis in momentum space. The bulk spectrum (blue dots) closes the gap at zero energy at two nodes, which are connected by dispersionless edge states (red). Numerical parameters are $k_{so} L_x \approx 4290$, $\mu =0$, $\Delta/E_{so} \approx 0.2$, $\phi=\pi$, $\Gamma/E_{so}\approx 0.2$, and $\Delta_Z/ E_{so} \approx 0.1$.}
	\label{WeylSpectrum}
\end{figure}

As discussed in the main text, in the regime $|\Delta_Z-\Delta|<\Gamma<|\Delta_Z+\Delta|$, the system is a 2D Weyl superconductor \cite{Tanaka2010,Sato2010,Sato2011,Schnyder2011,Meng2012,Wong2013,Deng2014,Schnyder2015}. The bulk spectrum is closed at two nodes around which the bulk spectrum is linear, {\it i.e.} the low-energy spectrum can be described by 2D Weyl cones. The position of the Weyl cones in momentum space is such that the line connecting the two cones is orthogonal to the direction of the Zeeman field. As already mentioned in the main text, in a Weyl superconductor (as well as in a Weyl semimetal) edge states only occur on edges at which the nodes of the two Weyl cones (being projected onto the edge) are not projected onto the same point. This means that the occurrence of edge states at the given edge in the 2D Weyl superconducting phase depends on the orientation ${\bf n}$ of the Zeeman field. In momentum space, the edge states appear on a line connecting the nodes of the Weyl cones, and since they are at zero energy, the corresponding edge states are flat (see Fig. \ref{WeylSpectrum}).

\section*{Appendix C: EDGE STATE WAVEFUNCTIONs IN HTSC PHASE}

In this section we provide more details on the calculation of the wavefunction of the helical edge states in the HTSC phase. The Hamiltonian density in the absence of a Zeeman field ($\Delta_Z=0$) [see Eqs. (1)-(3) in the main text] reads
\begin{equation}
\mathcal{H}_0(k_x,k_y)=\left[\eta_3 \frac{\hbar^2 k^2}{2m} + \tau_3 \alpha (k_y \sigma_1 - k_x \eta_3 \sigma_2) + \Gamma \tau_1 \eta_3 + \Delta \tau_3 \eta_2 \sigma_2\right]/2, \label{HamiltonianNambu1}
\end{equation}
where $\tau_i$, $\eta_i$, $\sigma_i$ are the Pauli matrices acting in layer, particle-hole, and spin space, respectively. Note that we set here $\phi=\pi$ for simplicity. This Hamiltonian density was also used to obtain the spectrum in Eq. (5) in the main text.

As discussed in the main text, we calculate the wavefunctions of the helical edge states that are exponentially localized at the $s=0$ edge. These modes are localized in the $y$ direction (located close to $y=0$) and propagate in the $x$ direction. In order to find the wavefunction, we assume the edge to be infinitely extended and focus on the states with $k_x=0$. Due to particle-hole and time-reversal symmetry it is clear that these states are at zero energy and twofold degenerate (Kramers pair). For simplicity, we work in the strong SOI limit where we can linearize the Hamiltonian density $\mathcal{H}_0(0,k_y)$ around two Fermi points ($k_{Fi}=0$, $k_{Fe}=2k_{so}$) \cite{Klinovaja2012},
\begin{equation}\label{HzeroKx}
\mathcal{H}_0(0,k_y) = \left[\eta_3 \frac{\hbar^2 k_y^2}{2m} + \tau_3 \sigma_1 \alpha k_y + \Gamma \tau_1 \eta_3 + \Delta \tau_3 \eta_2 \sigma_2\right]/2\, .
\end{equation}
In what follows we denote the wavefunctions (field operators) by $\tilde{\Phi}_{\sigma}$ ($\tilde{\psi}_{\sigma}$) when $\sigma = \uparrow,\downarrow$ refers to the spin projection onto the $x$ axis. The field operators can then be approximated as
\begin{align}
\tilde{\psi}_{1 \uparrow} &= \tilde{L}_{1 \uparrow}e^{-2i k_{so} y} + \tilde{R}_{1 \uparrow},\
\tilde{\psi}_{1 \downarrow} = \tilde{L}_{1 \downarrow} + \tilde{R}_{1 \downarrow} e^{2i k_{so} y} \nonumber \\
\tilde{\psi}_{\bar{1} \uparrow} &= \tilde{L}_{\bar{1} \uparrow} + \tilde{R}_{\bar{1} \uparrow}e^{2i k_{so}}, \  \tilde{\psi}_{\bar{1} \downarrow} = \tilde{L}_{\bar{1} \downarrow} e^{-2i k_{so} y} + \tilde{R}_{\bar{1} \downarrow}, 
\end{align}
where $\tilde{L}_{\tau \sigma}(y)$ [$\tilde{R}_{\tau \sigma}(y)$] are slowly varying fields on the scale of $k_{so}^{-1}$. The linearized Hamiltonian density is then given by
\begin{equation} \label{HamiltonianLinearized}
\tilde{\mathcal{H}}_{0,lin} = [\hbar v_F \hat{k} \rho_3 + \Gamma (\tau_1 \eta_3 \rho_1 - \tau_2 \eta_3 \sigma_3 \rho_2)/2 + \Delta \tau_3 \eta_2 \sigma_2 \rho_1]/2\, ,
\end{equation}
where $\hat{k}=-i \hbar \partial_y$ is the momentum operator and $v_F=\alpha/\hbar$ the Fermi velocity, and $\rho$ actin left/right mover space. 
By imposing vanishing boundary conditions, $\tilde{\Phi}_{\pm}(y=0) =0$, we find two-independent zero-energy solutions only if $\Gamma>\Delta$. The corresponding wavefunctions are written in the basis $\tilde{\Psi}=(\tilde{\psi}_{1 \uparrow, k_x=0}, \tilde{\psi}_{1 \downarrow, k_x=0}, \tilde{\psi}^{\dagger}_{1, \uparrow, k_x=0},  \tilde{\psi}^{\dagger}_{1 \downarrow,k_x=0},\tilde{\psi}_{\bar{1} \uparrow, k_x=0}, \tilde{\psi}_{\bar{1} \downarrow, k_x=0}, \tilde{\psi}^{\dagger}_{\bar{1} \uparrow, k_x=0},  \tilde{\psi}^{\dagger}_{\bar{1} \downarrow,k_x=0})$ as
\begin{align}\label{MF}
\tilde{\Phi}_+(y) &=\frac{1}{\sqrt{N}} \Big[f_1(y), g_{1}(y),f^*_{1}(y), g^*_{1}(y),f_{\bar{1}}(y), g_{\bar{1}}(y),f^*_{\bar{1}}(y), g^*_{\bar{1}}(y)\Big]^T, \nonumber \\
\tilde{\Phi}_-(y) &= \frac{1}{\sqrt{N}} \Big[g^*_{1}(y), -f^*_{1}(y), g_{1}(y), -f_{1}(y) ,g^*_{\bar{1}}(y), -f^*_{\bar{1}}(y), g_{\bar{1}}(y), -f_{\bar{1}}(y)\Big]^T,
\end{align}
with $f_1(y)=-i g_1^*(y)=e^{2i k_{so}y}e^{-y/\xi}+e^{-y/\xi_-}$ and $f_{\bar{1}}(y)=-i g_{\bar{1}}^*(y)=-i(e^{-2ik_{so}y} e^{-y/\xi}+e^{-y/\xi_-})$. These two solutions are Kramers partners and related by time-reversal symmetry described by the operator $\Theta=-i \sigma_2 \mathcal{K}$ with $\Theta \tilde{\Phi}_\pm= \pm \tilde{\Phi}_{\mp}$. Moreover, both $\tilde{\Phi}_+(y)$ and $\tilde{\Phi}_-(y)$ are extended Majorana wavefunctions (not to be confused with the MCSs that are localized at the corners) that are eigenstates of the particle-hole operator $\mathcal{P}=\eta_1 \mathcal{K}$, i.e., $\mathcal{P} \tilde{\Phi}_{\pm}= \tilde{\Phi}_{\pm}$. As can be seen above, the spin-up and spin-down components of the Majorana wavefunctions in Eq.~\eqref{MF} are not independent but related by $g_{\tau}(y)=-i f^*_{\tau}(y)$ \cite{Klinovaja2014,Yacoby}. This is ensured by the symmetry operator $ \mathcal{O}=\eta_2 \sigma_1$, which commutes with the particle-hole symmetry operator $[\mathcal{P},\mathcal{O}]=0$, and thus, $\mathcal{O} \tilde{\Phi}_{\pm}=\tilde{\Phi}_{\pm}$. The operator $\mathcal{O}$ anti-commutes with the Hamiltonian density in Eq. \eqref{HzeroKx}, $\{\mathcal{H}_0,\mathcal{O}\}=0$. Moreover, also the functions $f_1(y)$ and $f_{\bar{1}}(y)$ are not independent from each other, but related by $f_1(y)=-i f_{\bar{1}}^*(y)$. This is ensured by the symmetry operator $\mathcal{O}'=\tau_1 \eta_2$, which commutes with the previous two symmetries, $[\mathcal{P},\mathcal{O}']=[\mathcal{O},\mathcal{O}']=0$, and thus $\mathcal{O}' \tilde{\Phi}_\pm = \pm \tilde{\Phi}_\pm$. It is useful to take advantage of these hidden symmetries when calculating the various matrix elements in deriving the effective low-energy Hamiltonian in the following.

First we note that the spin-operator $\tilde{{\bf S}}$ in Nambu space is parametrized by the following Pauli matrices $\tilde{{\bf S}} \sim  (\eta_3 \sigma_3, \sigma_2, \eta_3 \sigma_1)$ (we recall that we take the $x$ axis as the spin quantization axis). In addition, this operator anti-commutes with the particle-hole operator, $\{\tilde{\bf S},\mathcal{P} \}=0$. Then, the symmetry of the Majorana wavefunctions leads to
\begin{equation}\label{Spin1}
\braket{\tilde{\Phi}_a|\tilde{\bf S}|\tilde{\Phi}_a}=\braket{\tilde{\Phi}_a|\mathcal{P}\tilde{\bf S}\mathcal{P}|\tilde{\Phi}_a} = -\braket{\tilde{\Phi}_a|\tilde{\bf S}|\tilde{\Phi}_a},
\end{equation}
for $a \in \{+,-\}$, and therefore $\braket{\tilde{\Phi}_a|\tilde{\bf S}|\tilde{\Phi}_a}=0$. The expectation values of spin projections to any axis is zero for Majorana states.
The time-reversal operator $\Theta$ and the operator $\mathcal{O}$, which ensures the structure of the Majorana wavefunctions $\tilde{\Phi}_\pm$, fix the values of the off-diagonal terms,
\begin{align}\label{Spin2}
&\braket{\tilde{\Phi}_+|\eta_3 \sigma_3|\tilde{\Phi}_-} = \braket{\tilde{\Phi}_+|\mathcal{P}\eta_3 \sigma_3 \Theta|\tilde{\Phi}_+}=\braket{\tilde{\Phi}_+|\eta_1 \mathcal{K}\eta_3 \sigma_3 (-i \sigma_2) \mathcal{K}|\tilde{\Phi}_+}= \braket{\tilde{\Phi}_+|\eta_1\eta_3 \sigma_3 (-i\sigma_2)|\tilde{\Phi}_+}=i\braket{\tilde{\Phi}_+|\eta_2 \sigma_1|\tilde{\Phi}_+}=i\braket{\tilde{\Phi}_+|\mathcal{O}|\tilde{\Phi}_+} \nonumber \\
&\hspace{2cm}= i\braket{\tilde{\Phi}_+|\tilde{\Phi}_+}=i, \nonumber \\
&\braket{\tilde{\Phi}_+|\sigma_2|\tilde{\Phi}_-}=\braket{\tilde{\Phi}_+|\mathcal{O}\sigma_2\mathcal{O}|\tilde{\Phi}_-}=-\braket{\tilde{\Phi}_+|\sigma_2|\tilde{\Phi}_-}=0\\
&\braket{\tilde{\Phi}_+|\eta_3 \sigma_1|\tilde{\Phi}_-}=\braket{\tilde{\Phi}_+|\mathcal{O}\eta_3 \sigma_1 \mathcal{O}|\tilde{\Phi}_-}=-\braket{\tilde{\Phi}_+|\eta_3 \sigma_1|\tilde{\Phi}_-}=0, \\
\end{align}
where we used the fact that $\{\mathcal{O},\eta_2\}=\{\mathcal{O},\eta_3 \sigma_1\}=0$.
In addition, one finds
\begin{align}\label{SC1}
&\braket{\tilde{\Phi}_+|(1-\tau_3)\eta_1 \sigma_2|\tilde{\Phi}_+} =\braket{\tilde{\Phi}_+|\eta_1 \sigma_2|\tilde{\Phi}_+}-\braket{\tilde{\Phi}_+|\tau_3\eta_1 \sigma_2|\tilde{\Phi}_+}=0, \text{ since}\nonumber \\
&\braket{\tilde{\Phi}_+|\eta_1 \sigma_2|\tilde{\Phi}_+}=\braket{\tilde{\Phi}_+|\mathcal{O}\eta_1 \sigma_2|\tilde{\Phi}_+}=\braket{\tilde{\Phi}_+|\eta_3 \sigma_3|\tilde{\Phi}_+}=0, \text{ and} \nonumber \\
&\braket{\tilde{\Phi}_+|\tau_3\eta_1 \sigma_2|\tilde{\Phi}_+}=\braket{\tilde{\Phi}_+|\mathcal{P}\tau_3\eta_1 \sigma_2|\tilde{\Phi}_+} =-\braket{\tilde{\Phi}_+|\tau_3\eta_1 \sigma_2|\tilde{\Phi}_+} =0,
\end{align}
where we used that $\{\mathcal{P}, \tau_3 \eta_1 \sigma_2 \}=0$. By analogy, one also finds $\braket{\tilde{\Phi}_-|(1-\tau_3)\eta_1 \sigma_2|\tilde{\Phi}_-} =0$. However, the off-diagonal term is given by
\begin{align}\label{SC2}
&\braket{\tilde{\Phi}_+|(1-\tau_3)\eta_1 \sigma_2|\tilde{\Phi}_-} =i, \text{ since} \nonumber \\
&\braket{\tilde{\Phi}_+|\eta_1 \sigma_2|\tilde{\Phi}_-} = \braket{\tilde{\Phi}_+|\eta_3 \sigma_3|\tilde{\Phi}_-} =i \text{ and} \nonumber \\
&\braket{\tilde{\Phi}_+|\tau_3\eta_1 \sigma_2|\tilde{\Phi}_-} = \braket{\tilde{\Phi}_+|\mathcal{P}\tau_3\eta_1 \sigma_2|\tilde{\Phi}_-} =- \braket{\tilde{\Phi}_+|\tau_3\eta_1 \sigma_2|\tilde{\Phi}_-} =0,
\end{align}
where in the second line we used Eq. \eqref{Spin2} and Eq. \eqref{SC1}. The last matrix element we are considering here is given by $\braket{\tilde{\Phi}_+|\tau_3 \eta_3 \sigma_2|\tilde{\Phi}_-}$. Using time-reversal and particle-hole symmetry operators, one can rewrite it as
\begin{align}\label{Skx1}
\braket{\tilde{\Phi}_+|\tau_3 \eta_3 \sigma_2|\tilde{\Phi}_-} = \braket{\tilde{\Phi}_+|\tau_3 \eta_3 \sigma_2(-i \sigma_2 \mathcal{K}) \eta_1 \mathcal{K}|\tilde{\Phi}_+} = -i\braket{\tilde{\Phi}_+|\tau_3 \eta_3 \eta_1 |\tilde{\Phi}_+} = -\braket{\tilde{\Phi}_+|\tau_3 \eta_2 |\tilde{\Phi}_+}.
\end{align}
This last expression can be shown to be zero by invoking the symmetry operator $\mathcal{O}'$,
\begin{align}\label{Skx2}
\braket{\tilde{\Phi}_+|\tau_3 \eta_2 |\tilde{\Phi}_+} = \braket{\tilde{\Phi}_+|\mathcal{O}'\tau_3 \eta_2 |\tilde{\Phi}_+}=\braket{\tilde{\Phi}_+|\tau_1 \eta_2\tau_3 \eta_2 |\tilde{\Phi}_+} = - \braket{\tilde{\Phi}_+|\tau_3 \eta_2 |\tilde{\Phi}_+} =0.
\end{align}
Importantly, the values of all these expectation values do not depend on the explicit form of $f_{\tau}(y)$.

The wavefunctions in the original spin basis, where the $z$ direction is the spin quantization axis, are given by $\Phi_{\pm }^{(0)}(y) = e^{-i \pi  \sigma_2/4} \tilde{\Phi}_{\pm}(y)$, which will be used in further calculations below. In order to avoid confusion we stress that the spin operator then takes the more familiar form ${\bf S}=\hbar (\eta_3 \sigma_1, \sigma_2, \eta_3 \sigma_3)/2$.

\section*{Appendix D: DERIVATION OF EFFECTIVE LOW-ENERGY HAMILTONIAN}

Using the explicit form of the wavefunctions obtained in the previous section, we derive the low-energy effective Hamiltonian on all four edges. As outlined in the main text, the Hamiltonian of the Rashba layers [see Eq. $(1)$ in the main text] is invariant under rotations $U_{\varphi}=e^{i \varphi \eta_3  J_z/\hbar}$, generated by the total angular momentum operator $J_z=L_z+S_z$. 
This allows us, knowing the wavefunctions at $s=0$ edge, to find the wavefunctions at the other three edges by applying $U_{\varphi}$ to $\Phi^{(0)}_{\pm}(y)$: $\Phi^{(s)}_\pm(y)=U_{\varphi_s} \Phi^{(0)}_\pm(y)$ with $\varphi_s \in \{0, \pi/2, \pi,3\pi/2 \}$.

We first derive the low-energy effective Hamiltonian for the helical edge states on the $s=0$ edge from the wavefunctions obtained in the previous section. To achieve this, we add various perturbation to $\mathcal{H}_0(0,k_y) $ defined in Eq. \eqref{HzeroKx}. We restrict ourselves to first-order perturbation theory applied in the low-energy subspace spanned by $ \Phi^{(0)}_\pm$, in which we focus only on linear terms in $\Delta_Z$, $\Delta$, and $k_x$ 

{\it Kinetic part.} The dispersion of the helical edge states close to zero-energy is linear, see Fig. 3 of the main text. This can be also shown explicitly by treating the linear term in $k_x$ in $\mathcal{H}_0(k_x,k_y)$ [see Eq. (\ref{HamiltonianNambu1})], $\mathcal{H}_{k_x}=-\alpha \tau_3 \eta_3 \sigma_2   k_x/2$ as a perturbation. For this term, the off-diagonal elements vanish by symmetry as was shown in Eq. \eqref{Skx1}-\eqref{Skx2}. However, for the diagonal elements these arguments do not apply: $\braket{\Phi^{(0)}_{\pm}|\mathcal{H}_{k_x}|\Phi^{(0)}_{\pm}}$ depends on the explicit form of the function $f_{1}(y)$. As a result, we obtain the effective low-energy Hamiltonian for helical edge states to be written as
\begin{equation}
\mathcal{H}^{(0)}_{kin} = \hbar v k_x \beta_3,
\end{equation}
where the Pauli matrix $\beta_i$ acts in the low-energy subspace and the effective velocity $v$ of the linearly dispersing edge states is given by	
\begin{equation}
v= v_F\frac{\Delta}{\Gamma}<v_F.
\end{equation}
We note that the Fermi velocity of the edge modes is the same at all four edges due to the rotation invariance of the system without applied magnetic fields.

{\it Superconducting gap at edge states for $\phi\neq \pi$.}
In the main text, we also discussed the case where the phase difference between the two layers is not exactly $\phi=\pi$, but slightly deviating from it, i.e., $\phi=\pi+\delta \phi$, where $\delta \phi \ll 1$ can be positive or negative. This leads to an extra factor in the superconducting pairing term in the $\tau=\bar{1}$ layer: $\Delta_{\bar{1}}= \Delta e^{i \phi}= \Delta e^{i(\pi + \delta \phi)} \approx - \Delta (1+i \delta \phi)$.
The deviations from the $\pi$-junction is given by $\mathcal{H}_{\delta \phi} = - \delta \phi \Delta (1-\tau_3)\eta_1 \sigma_2/4$. In Eqs. \eqref{SC1} and \eqref{SC2}, we showed that, due to the system symmetry, only the off-diagonal matrix elements $\braket{\Phi^{(0)}_{\pm}|\mathcal{H}_{\delta \phi}|\Phi^{(0)}_{\mp}}$ are non-zero. This leads us to
\begin{equation}
\mathcal{H}^{(0)}_{sc} = -\Delta \delta \phi \beta_2/2.
\end{equation}
Again, due to the rotation invariance of the superconducting part of the Hamiltonian, this term opens a gap of the same size at all edges.

{\it Zeeman term.} For a weak Zeeman field $\Delta_Z\ll \Gamma - \Delta$, the corresponding term opening the gap in the edge state spectrum is given by $\mathcal{H}_Z=\Delta_Z [\cos (\theta) \eta_3 \sigma_1 + \sin( \theta) \sigma_2]/2$. Using Eqs. \eqref{Spin1} and \eqref{Spin2}, one notices that in the low-energy subspace only the term proportional to $\eta_3 \sigma_1$ is important, and only its off-diagonal matrix elements are non-zero. This leads us to the effective Zeeman term acting in the low-energy subspace, 
\begin{equation}
\mathcal{H}^{(0)}_Z = - \Delta_Z \cos (\theta) \beta_2.
\end{equation}
The form of the effective Zeeman term in the Hamiltonian for the remaining edges can be simply obtained by calculating the matrix elements $\braket{\Phi^{(s)}_a|\mathcal{H}_Z|\Phi^{(s)}_b}$, where we make use of the symmetry operator $U_{\varphi_s}$,
\begin{align}
&\left<\Phi^{(s)}_{a}|  {\cal H}_Z |\Phi^{(s)}_{b}\right>=\left<\Phi^{(0)}_{a}| U^{\dagger}_{\varphi_s} {\cal H}_Z U_{\varphi_s}|\Phi^{(0)}_{b}\right>=-\Delta_Z\cos(\theta-\varphi_s) \beta_2,\\
&U^{\dagger}_{\varphi_s} {\cal H}_Z U_{\varphi_s} =\Delta_Z[\cos(\theta-\varphi_s) \ \eta_3 \sigma_1 + \sin(\theta-\varphi_s) \sigma_2]/2.
\end{align}
Thus, the gap opened by the Zeeman field is different at different edges and is given by $\Delta_Z \cos(\theta- \varphi_s)$.
Collecting all terms together, we arrive at Eq. (6) of the main text:
\begin{align}
&\mathcal{H}^{(s)}_{eff}= \hbar v k^{(s)} \beta_3-\Delta_s \beta_2,\\
& \Delta_s= \Delta \delta \phi/2+\Delta_Z \cos(\theta- \varphi_s).
\end{align}
By examining the sign of the gap $ \Delta_s$, one can identify potential domain walls and determine the location of zero-energy MCSs.

\section{Appendix E: Stability of Majorana corner states against disorder}

\begin{figure}[t]
	\centering
	\includegraphics[width=\textwidth]{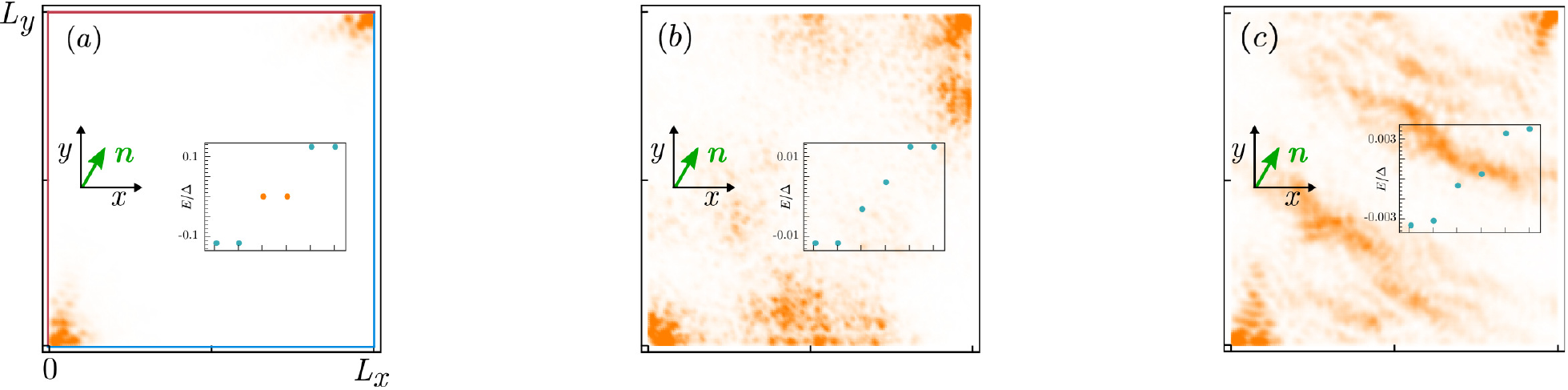}
	\caption{(a) The SOTSC phase is stable against fluctuations in chemical potential (potential disorder) and in Zeeman field (magnetic disorder). Here, we choose $\overline{\sigma}_{\mu}/E_{so} \approx 0.4$ and $\overline{\sigma}_Z/E_{so} \approx 0.1$. (b) If the static potential disorder strength is as large as $\overline{\sigma}_{\mu}/E_{so} \approx 1.6$, the bulk gap closes and, consequently, the MCSs disappear. (c) The same happens if the magnetic disorder strength is as large as $\overline{\sigma}_Z/E_{so} \approx 0.4$ . Numerical parameters are the same as in Fig. 4a in the main text: $k_{so}L_x=k_{so}L_y=430$, $\mu=0$, $\Delta/E_{so}=0.2$, $\Delta_Z/E_{so}=0.15$, $\Gamma/E_{so}=0.6$, $\theta=\pi/3$, and $\delta \phi=0$ (corresponding to $N_x=N_y=150$, $\mu=0$, $\Delta/t=0.06$, $\Delta_Z/t=0.05$, $\Gamma/t=0.17$, $\theta= \pi/3$, and $\delta \phi=0$ in the tight-binding model).}
	\label{Disorder}
\end{figure}

In this section, we show numerically that the SOTSC phase is stable against fluctuations of the local chemical potential (potential disorder) and of Zeeman field (magnetic disorder), see Fig. \ref{Disorder}.

{\it Potential disorder.}
In the considerations above we assumed the chemical potential to be fixed at the SOI energy, {\it i.e.} $\mu=0$. In the presence of scalar impurities the local potential can fluctuate around this value, which can be taken into account by introducing local fluctuations $\delta \mu_{ij}$ at lattice site $(i,j)$ that follow a normal distribution with zero mean value, {\it i.e.} $\braket{\delta \mu_{ij}}=0$. The strength of the potential fluctuations is characterized by the standard deviation $\overline{\sigma}_\mu=\sqrt{\braket{\delta \mu_{ij}^2}}$. 

{\it Magnetic disorder.}
Magnetic impurities create local fluctuations in the effective Zeeman field. As for potential disorder, this can be taken into account by introducing the local deviation from the Zeeman field ${\bf h}_{ij}= (h_{x,ij},h_{y,ij},h_{z,ij})$, where each component separately follows a normal distribution with the corresponding mean values, {\it i.e.} $\braket{h_{x,ij}}= \Delta_Z \cos \theta$, $\braket{h_{x,ij}}= \Delta_Z \sin \theta$, $\braket{h_{z,ij}}= 0$, and the standard deviation $\overline{\sigma}_{Z}=\sqrt{\braket{{\bf h}_{ij}^2}-\braket{{\bf h}_{ij}}^2}$. 	We find that the SOTSC phase is robust against strong potential disorder characterized by $\overline{\sigma}_\mu$ substantially exceeding the bulk gap. However, if disorder is too strong, the bulk gap closes and the system becomes effectively gapless. As a result, the MCSs disappear.
In addition, we note that the coupling between the parent $s$-wave superconductors and the Rashba layers should be weak to avoid metallization effects \cite{Reeg:2017_3,Reeg:2018,Reeg:2018_2,Antipov:2018} observed in one-dimensional Rashba nanowires in the strong coupling regime \cite{Chang:2015,Gazibegovic:2017,Kjaergaard:2016,Shabani:2016}.

\bibliographystyle{unsrt}

\begin{thebibliography}{}
	
	\bibitem{Kitaev2001}
	A. Y. Kitaev, Phys.-Usp. {\bf 44}, 131 (2001).
	\bibitem{DasSarma2010}
	R. M. Lutchyn, J. D. Sau, and S. D. Sarma, Phys. Rev. Lett. {\bf 105}, 177002 (2010).
	\bibitem{Oreg2010}
	Y. Oreg, R. Refael, and F. von Oppen, Phys. Rev. Lett. {\bf 105}, 177002 (2010).
	\bibitem{Alicea2010}
	J. Alicea, Phys. Rev. B {\bf 81}, 125318 (2010).
	\bibitem{KaneMele20051}
	C. L. Kane and E. J. Mele, Phys. Rev. Lett. {\bf 95}, 226801 (2005).
	\bibitem{KaneMele20052}
	C. L. Kane and E. J. Mele, Phys. Rev. Lett. {\bf 95}, 146802 (2005).
	\bibitem{Wu2006}
	C. Wu, B. A. Bernevig, and S. Zhang, Phys. Rev. Lett. {\bf 96}, 106401 (2006).
	\bibitem{Bernevig2006}
	B. A. Bernevig, T. L. Hughes, and S. Zhang, Science {\bf 314}, 5806 (2006).
	\bibitem{FuKane2007}
	L. Fu, C. L. Kane, and E. J. Mele, Phys. Rev. Lett. \textbf{98}, 106803 (2007).
	\bibitem{Moore2007}
	J. E. Moore and L. Balents, Phys. Rev. B \textbf{75}, 121306(R) (2007).
	\bibitem{Koenig2007}
	M. K\"onig, S. Wiedmann, C. Br\"une, A. Roth, H. Buhmann, L. W. Molenkamp, X. Qi, and S. Zhang, Science {\bf 318}, 5851 (2007).
	\bibitem{Hsieh2008}
	D. Hsieh, D. Qian, L. Wray, Y. Xia, Y. S. Hor, R. J. Cava, and M. Z. Hasan, Nature \textbf{452}, 970 (2008).
	\bibitem{Hsieh2009}
	D. Hsieh, Y. Xia, L. Wray, D. Qian, A. Pal, J. H. Dil, J. Osterwalder, F. Meier, G. Bihlmayer, C. L. Kane, Y. S. Hor, R.
	J. Cava, and M. Z. Hasan, Science \textbf{323}, 919 (2009).
	\bibitem{HasanKane2010}
	M. Z. Hasan and C. L. Kane, Rev. Mod. Phys. {\bf 82}, 3045 (2010).
	\bibitem{Zhang2011}
	X.-L. Qi and S.-C. Zhang, Rev. Mod. Phys. {\bf 83}, 1057 (2011).
	\bibitem{Ando2015}
	Y. Ando and L. Fu, Annu. Rev. Condens. Matter Phys. {\bf 6}, 361 (2015).
	\bibitem{Stern2016}
	A. Stern, Annu. Rev. Condens. Matter Phys. {\bf 7}, 349 (2016).
	\bibitem{Sato2017}
	M. Sato and Y. Ando, Rep. Prog. Phys. {\bf 80}, 076501 (2017).
	\bibitem{Potter2011}
	A. C. Potter and P. A. Lee, Phys. Rev. B {\bf 83}, 094525 (2011).
	\bibitem{Hasan2011}
	M. Z. Hasan and J. E. Moore, Annu. Rev. Condens. Matter {\bf 2}, 55 (2011).
	\bibitem{Chevallier2012}
	D. Chevallier, D. Sticlet, P. Simon, and C. Bena, Phys. Rev. B {\bf 85}, 235307 (2012).
	\bibitem{Klinovaja2012_3} J. Klinovaja, S. Gangadharaiah, and D. Loss, Phys. Rev. Lett. {\bf 108}, 196804 (2012).
	\bibitem{Sticlet2012} D. Sticlet, C. Bena, and P. Simon, Phys. Rev. Lett. {\bf 108}, 096802 (2012).
	\bibitem{Klinovaja2012_2} J. Klinovaja, P. Stano, and D. Loss, Phys. Rev. Lett. {\bf 109}, 236801 (2012).
	\bibitem{Halperin2012} B. I. Halperin, Y. Oreg, A. Stern, G. Refael, J. Alicea, and F. von Oppen, Phys. Rev. B {\bf 85}, 144501 (2012).
	\bibitem{NadjPerge2013}
	S. Nadj-Perge, I. K. Drozdov, B. A. Bernevig, and A. Yazdani, Phys. Rev. B {\bf 88}, 020407 (R) (2013).
	\bibitem{Klinovaja2013} J. Klinovaja, P. Stano, A. Yazdani, and D. Loss, Phys. Rev. Lett. {\bf 111}, 186805 (2013).
	\bibitem{Braunecker2013} B. Braunecker and P. Simon, Phys. Rev. Lett. {\bf 111}, 147202 (2013).
	\bibitem{Vazifeh2013} M. Vazifeh and M. Franz, Phys. Rev. Lett. {\bf 111}, 206802 (2013).
	\bibitem{Weithofer2014}
	L. Weithofer, P. Recher, and T. L. Schmidt, Phys. Rev. B {\bf 90}, 205416 (2014).
	\bibitem{Manousakis2017}
	J. Manousakis, A. Altland, D. Bagrets, R. Egger, and Y. Ando, Phys. Rev. B {\bf 95}, 165424 (2017).
	\bibitem{Das2012}
	A. Das, Y. Ronen, Y. Most, Y. Oreg, M. Heiblum, and H, Sh.rikman, Nature {\bf 8}, 887–895 (2012).
	\bibitem{Mourik2012}
	V. Mourik, K. Zuo, S. M. Frolov, S. R. Plissard, E. P. A. M. Bakkers, and L. P. Kouwenhoven, Science {\bf 336}, 6084 (2012).
	\bibitem{Deng2012}
	M. T. Deng, C. L. Yu, G. Y. Huang, M. Larsson, P. Caroff, and H. Q. Xu, Nano Lett. {\bf 12}, 6414 (2012).
	\bibitem{Finkc2013}
	A. D. K. Finck, D. J. Van Harlingen, P. K. Mohseni, K. Jung, and X. Li, Phys. Rev. Lett. {\bf 110}, 126406 (2013).
	\bibitem{Churchill2013}
	H. O. H. Churchill, V. Fatemi, K. Grove-Rasmussen, M. T. Deng, P. Caroff, H. Q. Xu, and C. M. Marcus, Phys. Rev. B {\bf 87}, 241401(R) (2013).
	\bibitem{Albrecht2016}
	S. M. Albrecht, A. P. Higginbotham, M. Madsen, F. Kuemmeth, T. S. Jespersen, J. Nygard, P. Krogstrup, and C. M. Marcus, Nature {\bf 531}, 206 (2016).
	\bibitem{Zhan2018}
	H. Zhan {\it et al.}, Nature {\bf 556}, 74 (2018).
	\bibitem{Benalcazar}
	W. A. Benalcazar, J. C. Teo, and T. L. Hughes, Phys. Rev. B {\bf 89}, 224503 (2014).
	\bibitem{Benalcazar2017}
	W. A. Benalcazar, B. A. Bernevig, and T. L. Hughes, Science {\bf 357}, 61 (2017).
	\bibitem{Benalcazar20172}
	W. A. Benalcazar, B. A. Bernevig, and T. L. Hughes, Phys. Rev. B {\bf 96}, 245115 (2017).
	\bibitem{song2017}
	Z. Song, Z. Fang, and C. Fang, Phys. Rev. Lett. {\bf 119}, 246402 (2017).
	\bibitem{Peng2017}
	Y. Peng, Y. Bao, and F. von Oppen, Phys. Rev. B {\bf 95}, 235143 (2017).
	\bibitem{Imhof2017}
	S. Imhof, C. Berger, F. Bayer, H. Brehm, L. Molenkamp, T. Kiessling, F. Schindler, C. H. Lee, M. Greiter, T. Neupert, and R. Thomale, arix:1708.03647.
	\bibitem{Geier2018}
	M. Geier, L. Trifunovic, M. Hoskam, and P. W. Brouwer, Phys. Rev. B {\bf 97}, 205135 (2018).
	\bibitem{Schindler2018}
	F. Schindler, A. M. Cook, M. G. Verginory, Z. Wang, S. S. P. Parking, B. A. Bernevig, and T. Neupert, Science Adv. {\bf 4}, 6 (2018).
	\bibitem{Hsu2018}
	C.-H. Hsu, P. Stano, J. Klinovaja, and D. Loss, arixv: 1805.12146.
	\bibitem{Ezawa2018}
	M. Ezawa, Phys. Rev. B {\b 97}, 155305 (2018).
	\bibitem{Ezawa20182}
	M. Ezawa, arxiv: 1807.10932.
	\bibitem{Ezawa20183}
	M. Ezawa, Phys. Rev. Lett. {\bf 121}, 116801 (2018).
	\bibitem{Zhu2018}
	X. Zhu, Phys. Rev. B {\bf 97}, 205134 (2018).
	\bibitem{Wang2018}
	Q. Wang, C.-C. Liu, Y.-M. Lu, and F. Zhang, arxiv:1804.04711.
	\bibitem{Yan2018}
	Z. Yan, F. Song, and Z. Wang, Phys. Rev. Lett. {\bf 121}, 096803 (2018).
	\bibitem{Liu2018}
	T. Liu, J. J. He, and F. Nori, arXiv:1806.07002.
	\bibitem{Zhang2018}
	X. Zhang, H.-X. Wang, Z.-K. Lin, Z. Tian, B. Xie, M.-H. Lu, Y.-F. Chen, and J.-H. Jian, arxiv:1806.10028.
	\bibitem{Wang20182}
	Q. Wang, D. Wand, and Q.-H. Wang, arxiv:1808.01085.
	\bibitem{Yacoby2018}
	H. Ren, F. Pientka, S. Hart, A. Pierce, M. Kosowsky, L. Lunczer, R. Schlereth, B. Scharf, E. M. Hankiewicz, L. W. Molenkamp, B. I. Halperin, and A. Yacoby, arXiv:1809.03076.
	\bibitem{Nichele2018}
	A. Fornieri, A. M. Whiticar, F. Setiawan, E. P. Martin, A. C. C. Drachmann, A. Keselman, S. Gronin, C. Thomas, T. Wang, R. Kallaher, G. C. Gardner, E. Berg, M.J. Manfra, A. Stern, C. M. Marcus, and F. Nichele, arXiv:1809.03037.
	\bibitem{Schrade2015}
	C. Schrade, A. A. Zyuzin, J. Klinovaja, and D. Loss, Phys. Rev. Lett. {\bf 115}, 237001 (2015).
	\bibitem{Buzdin} A. I. Buzdin, L. N. Bulaevskii, and S.V. Panyukov, Pis’ma Zh. Eksp. Teor. Fiz. {\bf 35}, 147 (1982) [JETP Lett. {\bf 35}, 178 (1982)].
	\bibitem{Ryazanov} V.V. Ryazanov, V.A. Oboznov, A. Yu. Rusanov, A.V. Veretennikov, A.A. Golubov, and J. Aarts, Phys. Rev. Lett. {\bf 86}, 2427 (2001).
	\bibitem{Spivak} B. I. Spivak and S. A. Kivelson, Phys. Rev. B {\bf 43}, 3740 (1991).
	\bibitem{Dam} J. A. van Dam, Y.V. Nazarov, E. P.A.M. Bakkers, and L. P. Kouwenhoven, Nature (London) {\bf 442}, 667 (2006).
	\bibitem{foot} We note that our result stays valid also if the two SOI vectors are slightly misaligned. The topological transition is determined by the gap closing at $k=0$, where the SOI terms do not play a role, under the condition that the bulk spectrum stays gapped at all other values of $k$.
		\bibitem{Keselman2013}
	A. Keselman, L. Fu, and E. Berg, Phys. Rev. Lett. {\bf 111}, 116402 (2013).
	\bibitem{Pientka2017}
	F. Pientka, A. Keselman, E. Berg, A. Yacoby, A. Stern, and B. I. Halperin, Phys. Rev. X {\bf 7}, 021032 (2017).
	\bibitem{SM}
	We refer to the Supplemental Material for the explicit form of the discretized Hamiltonians, discussion of the 2D Weyl SC phase, details on the calculation of the wavefunctions in HTSC phase, the derivation of the low-energy effective Hamiltonian as well as the treatment of disorder.
	\bibitem{Ryu2010}
	S. Ryu, A. P. Schnyder, A. Furusaki, and A. W. W. Ludwig, New J. Phys. {\bf 12}, 065010 (2010).
	\bibitem{Tanaka20101}
	Y. Tanaka, Y. Mizuno, T. Yokoyama, K. Yada, and M. Sato, Phys. Rev. Lett. {\bf 105}, 097002 (2010).
	\bibitem{Sato20101} 
	M. Sato and S. Fujimoto, Phys. Rev. Lett. {\bf 105}, 217001 (2010).
	\bibitem{Sato20111}
	M. Sato, Y. Tanaka, K. Yada, and T. Yokoyama, Phys. Rev. B {\bf 83}, 224511 (2011).
	\bibitem{Schnyder20111} 
	A. P. Schnyder and S. Ryu, Phys. Rev. B {\bf 84}, 060504(R) (2011).
	\bibitem{Meng20121}
	T. Meng and L. Balents, Phys. Rev. B {\bf 86}, 054504 (2012).
	\bibitem{Wong20131}
	C. L. M. Wong, J. Liu, K. T. Law, and P. A. Lee, Phys. Rev. B {\bf 88}, 060504(R) (2013).
	\bibitem{Deng20141}
	S. Deng, G. Ortiz, A. Poudel, and L. Viola, Phys. Rev. B {\bf 89}, 140507(R) (2014).
	\bibitem{Schnyder20151}
	A. P. Schnyder and P. M. R. Brydon, J. Phys.: Condens. Matter {\bf 27}, 243201 (2015).
	\bibitem{Hao20171}
	L. Hao and C. S. Ting, Phys. Rev. B {\bf 95}, 064513 (2017).
	\bibitem{Volpez2018}
	Y. Volpez, D. Loss, and J. Klinovaja, Phys. Rev. B {\bf 97}, 195421 (2018).
	\bibitem{Jackiw1976}
	R. Jackiw and C. Rebbi, Phys. Rev. D 13, 3398 (1976).
	\bibitem{Jackiw1981}
	R. Jackiw and J. Schrieffer, Nucl. Phys. B 190, 253 (1981).
	\bibitem{BlackSchaffer2016}
	F. Parhizgar and A. Black-Schaffer, Sci. Rep. {\bf 7}, 9817 (2017).
\end{thebibliography}

\begin{thebibliography}{}
	
	\bibitem{Diego} D. Rainis, L. Trifunovic, J. Klinovaja, and D. Loss, Phys. Rev. B {\bf 87}, 024515 (2013).
	
	\bibitem{SOI} J. Klinovaja and D. Loss, Eur. Phys. J. B {\bf 88}, 62 (2015).
	
	\bibitem{Tanaka2010}
	Y. Tanaka, Y. Mizuno, T. Yokoyama, K. Yada, and M. Sato, Phys. Rev. Lett. {\bf 105}, 097002 (2010).
	\bibitem{Sato2010} 
	M. Sato and S. Fujimoto, Phys. Rev. Lett. {\bf 105}, 217001 (2010).
	\bibitem{Sato2011}
	M. Sato, Y. Tanaka, K. Yada, and T. Yokoyama, Phys. Rev. B {\bf 83}, 224511 (2011).
	\bibitem{Schnyder2011} 
	A. P. Schnyder and S. Ryu, Phys. Rev. B {\bf 84}, 060504(R) (2011).
	\bibitem{Meng2012}
	T. Meng and L. Balents, Phys. Rev. B {\bf 86}, 054504 (2012).
	\bibitem{Wong2013}
	C. L. M. Wong, J. Liu, K. T. Law, and P. A. Lee, Phys. Rev. B {\bf 88}, 060504(R) (2013).
	\bibitem{Deng2014}
	S. Deng, G. Ortiz, A. Poudel, and L. Viola, Phys. Rev. B {\bf 89}, 140507(R) (2014).
	\bibitem{Schnyder2015}
	A. P. Schnyder and P. M. R. Brydon, J. Phys.: Condens. Matter {\bf 27}, 243201 (2015).
	
	\bibitem{Klinovaja2012}
	J. Klinovaja and D. Loss, Phys. Rev. B {\bf 86}, 085408 (2012).
	
	\bibitem{Klinovaja2014}
	J. Klinovaja and D. Loss, Phys. Rev. B {\bf 90}, 045118 (2014).
	
	\bibitem{Yacoby} J. Klinovaja, A. Yacoby, and D. Loss,
	Phys. Rev. B {\bf 90}, 155447 (2014).
	
	\bibitem{Reeg:2017_3}{C. Reeg, D. Loss, and J. Klinovaja, Phys. Rev. B {\bf 96}, 125426 (2017).}
	\bibitem{Reeg:2018}{C. Reeg, D. Loss, and J. Klinovaja, Phys. Rev. B {\bf 97}, 165425 (2018).}
	\bibitem{Reeg:2018_2}{C. Reeg, D. Loss, and J. Klinovaja, Beilstein J. Nan- otechnol. {\bf 9}, 1263 (2018).}
	\bibitem{Antipov:2018}{A. E. Antipov, A. Bargerbos, G. W. Winkler, B. Bauer,
		E. Rossi, and R. M. Lutchyn, Phys. Rev. X {\bf 8}, 031041 (2018).}
	\bibitem{Chang:2015}{W. Chang, S. M. Albrecht, T. S. Jespersen, F. Kuemmeth, P. Krogstrup, J. Nyg\aa rd, and C. M. Marcus, Nat. Nano. {\bf 10}, 232 (2015).}
	\bibitem{Gazibegovic:2017}{S. Gazibegovic, D. Car, H. Zhang, S. C. Balk, J. A. Logan, M. W. A. de Moor, M. C. Cassidy, R. Schmits, D. Xu, G. Wang, P. Krogstrup, R. L. M. Op het Veld, K. Zuo, Y. Vos, J. Shen, D. Bouman, B. Shojaei, D. Pen- nachio, J. S. Lee, P. J. van Veldhoven, S. Koelling, M. A. Verheijen, L. P. Kouwenhoven, C. J. Palmstr\o m, and E. P. A. M. Bakkers, Nature {\bf 548}, 434 (2017).}
	\bibitem{Kjaergaard:2016}{M. Kjaergaard, F. Nichele, H. J. Suominen, M. P. Nowak, M. Wimmer, A. R. Akhmerov, J. A. Folk, K. Flensberg, J. Shabani, C. J. Palmstr\o m, and C. M. Marcus, Nat. Commun. {\bf 7}, 12841 (2016).}
	\bibitem{Shabani:2016}{J. Shabani, M. Kjaergaard, H. J. Suominen, Y. Kim, F. Nichele, K. Pakrouski, T. Stankevic, R. M. Lutchyn, P. Krogstrup, R. Feidenhans'l, S. Kraemer, C. Nayak, M. Troyer, C. M. Marcus, and C. J. Palmstr\o m, Phys. Rev. B {\bf 93}, 155402 (2016).}
\end{thebibliography}

 \end{document}